\documentclass[aps,prl,twocolumn,floatfix,superscriptaddress]{revtex4}

\usepackage{amssymb,amsmath,amstext}                
\usepackage{graphicx}                                               
\usepackage{color}                                                     
\usepackage{bm}                                                        
\usepackage{appendix}                                              
\usepackage[utf8]{inputenc}
\usepackage{bbold}
\usepackage{bbm}
\usepackage{ulem}
\usepackage{latexsym}
\usepackage[colorlinks=true,citecolor=blue,linkcolor=magenta]{hyperref}

\newcommand{\vect}[1]{\mathbf{#1}}

\def\be{\begin{equation}}
\def\ee{\end{equation}}
\def\bea{\begin{eqnarray}}
\def\eea{\end{eqnarray}}
\def\ra{\rangle}
\def\la{\langle}
\def\bi{\begin{itemize}}
\def\ei{\end{itemize}}

\definecolor{dgreen} {RGB}{78,138,21}

\begin{document} 

\title{Time crystal platform: from       
quasi-crystal structures in time to systems with exotic interactions}

\author{Krzysztof Giergiel} 
\affiliation{
Instytut Fizyki imienia Mariana Smoluchowskiego, 
Uniwersytet Jagiello\'nski, ulica Profesora Stanis\l{}awa \L{}ojasiewicza 11, PL-30-348 Krak\'ow, Poland}
\author{Artur Miroszewski}
\affiliation{
Instytut Fizyki imienia Mariana Smoluchowskiego, 
Uniwersytet Jagiello\'nski, ulica Profesora Stanis\l{}awa \L{}ojasiewicza 11, PL-30-348 Krak\'ow, Poland}
\affiliation{National Centre for Nuclear Research, ul.Ho\.za 69, PL-00-681 Warsaw, Poland}
\author{Krzysztof Sacha} 
\affiliation{
Instytut Fizyki imienia Mariana Smoluchowskiego, 
Uniwersytet Jagiello\'nski, ulica Profesora Stanis\l{}awa \L{}ojasiewicza 11, PL-30-348 Krak\'ow, Poland}
\affiliation{Mark Kac Complex Systems Research Center, Uniwersytet Jagiello\'nski, ulica Profesora Stanis\l{}awa \L{}ojasiewicza 11, PL-30-348 Krak\'ow, Poland
}

\begin{abstract}
Time crystals are quantum many-body systems which, due to interactions between particles, are able to spontaneously self-organize their motion in a periodic way in time by analogy with the formation of crystalline structures in space in condensed matter physics. In solid state physics properties of space crystals are often investigated with the help of external potentials that are spatially periodic and reflect various crystalline structures. A similar approach can be applied for time crystals, as periodically driven systems constitute counterparts of spatially periodic systems, but in the time domain. Here we show that condensed matter problems ranging from single particles in potentials of quasi-crystal structure to many-body systems with exotic long-range interactions can be realized in the time domain with an appropriate periodic driving. Moreover, it is possible to create molecules where atoms are bound together due to destructive interference if the atomic scattering length is modulated in time.
\end{abstract}
\date{\today}

\maketitle

Although crystals have been known for years, time crystals sound more like science fiction than a serious scientific concept. In 2012 Frank Wilczek initiated new research area by suggesting that periodic structures in time can be formed spontaneously by a quantum many-body system \cite{Wilczek2012}. While the original Wilczek idea could not be realized because it was based on a system in the ground state \cite{Bruno2013b,Watanabe2015,Syrwid2017,Iemini2017}, it turned out that spontaneous breaking of discrete time translation symmetry and self-re-organization of motion of a periodically driven quantum many-body system was possible \cite{Sacha2015}. This phenomenon was dubbed "discrete time crystals" \cite{Khemani16,ElseFTC,Yao2017} and it was already realized experimentally \cite{Zhang2017,Choi2017}, for review see \cite{Sacha2017rev}.   

Apart from the realization of spontaneous breaking of discrete time translation symmetry, periodically driven quantum systems can be also employed to model crystalline structures in time in a similar way as external time-independent spatially periodic potentials allow one to model space crystals \cite{Guo2013,Sacha15a}. It should be stressed that driven systems with crystalline properties in time do not require external spatially periodic potentials. Crystal structures in time emerge due to periodic driving provided it is resonant with the unperturbed motion of a system. It is possible to investigate Anderson localization \cite{MuellerDelande:Houches:2009} in the time domain \cite{Sacha15a,sacha16,Giergiel2017,delande17} or many-body localization caused by temporal disorder \cite{Mierzejewski2017}. In the following we show that proper manipulation of higher temporal harmonics of a periodic perturbation is a perfect tool to engineer a wide class of condensed matter systems in the time domain including many-body systems with exotic interactions. Moreover, it is possible to create molecules where atoms are bound together via disordered potentials.

Let us begin with a classical single particle system in one-dimension (1D) described by the Hamiltonian $H_0(x,p)$. If the motion of a particle is bounded it is convenient to perform a canonical transformation to the so-called action-angle variables \cite{Lichtenberg1992}. Then, $H_0=H_0(I)$ where the momentum (action) $I$ is a constant of motion and the canonically conjugate angle $\theta$ changes linearly with time, i.e. $\theta(t)=\Omega t+\theta(0)$ where $\Omega(I)=\frac{dH_0(I)}{dI}$ is a frequency of periodic evolution of a particle. Assume we turn on a periodic driving of the form $H_1=\lambda h(x)f(t)$ where $f(t+2\pi/\omega)=f(t)=\sum_{k}f_{k}e^{ik\omega t}$ and $\lambda$ determines the strength of the driving. The spatial part of $H_1$ can be expanded in a Fourier series $h(x)=\sum_n h_n(I)e^{in\theta}$. A particle will be resonantly driven if the period of its unperturbed motion is equal to an integer multiple of the driving period, i.e. $\omega=s\Omega(I_0)$ where $s$ is an integer number and $I_0$ is a resonant value of the action. In order to analyze motion of a particle in the vicinity of a resonant trajectory, i.e. for $I\approx I_0$, it is convenient to switch to the moving frame, $\Theta=\theta-\frac{\omega}{s}t$, and apply the secular approximation \cite{Lichtenberg1992}. It results in the effective time-independent Hamiltonian, $H_{\rm eff}=\frac{P^2}{2m_{\rm eff}}+\lambda V_{\rm eff}(\Theta)$, where $P=I-I_0$, the effective mass $\frac{1}{m_{\rm eff}}=\frac{d^2H_0(I_0)}{dI_0^2}$ and the potential $V_{\rm eff}(\Theta)=\sum_n h_{ns}(I_0)f_{-n}e^{ins\Theta}$. If the second order corrections are negligible, which can be easily monitored \cite{Lichtenberg1992,supplement}, $H_{\rm eff}$ provides an exact description of particle motion in the vicinity of a resonant trajectory. 

The effective Hamiltonian, $H_{\rm eff}$, indicates that a resonantly driven system behaves like a particle on a ring, i.e. $0<\Theta\le 2\pi$, with a certain effective mass and in the presence of a time-independent effective potential $V_{\rm eff}(\Theta)$. If there are many non-zero Fourier components of $h(x)$, a proper choice of temporal Fourier components of $f(t)$ allows one to create a practically arbitrary effective potential. Indeed, any potential on a ring can be expanded in a series $V_{\rm eff}(\Theta)=\sum_nd_ne^{in\Theta}$ and in order to realize it we can choose the fundamental 1:1 resonance ($s=1$) and periodic driving with the Fourier components $f_{-n}=\frac{d_n}{h_{n}(I_0)}$. If $s>1$, a potential energy structure is duplicated $s$ times. For $s\gg 1$, $V_{\rm eff}(\Theta)$ allows one to reproduce condensed matter problems where a particle can move in a potential with $s$ identical wells of arbitrary shape and with periodic boundary conditions.

Before we illustrate our idea with an example we have to address two issues. Firstly, so far our approach was classical but we would like to deal with quantum systems. In order to obtain a quantum description one can either perform quantization of the effective Hamiltonian, i.e. $(P,\Theta)\rightarrow(\hat P,\hat\Theta)$, or apply a quantum version of the secular approximation from the very beginning \cite{Berman1977}. Both approaches lead to the same results. Eigenstates of the effective Hamiltonian in the moving frame correspond to time-periodic Floquet eigenstates of the original Floquet Hamiltonian, $H_F=H_0+H_1-i\hbar\partial_t$, in the laboratory frame \cite{Buchleitner2002,supplement}. Second issue: what is the relation of the class of problems we consider with time crystals? Space crystals are related to periodic arrangement of particles in space. If we take a snapshot of a space crystal at some moment in time ($t=$const.), then we can observe a crystalline structure in space. Switching to time crystals the role of time and space is exchanged. We fix position in the configuration space ($x=$const.), i.e. we choose location for the detector, and ask if the probability of clicking of a detector behaves periodically in time. We have shown that in the frame moving along a classical resonant orbit, $\Theta=\theta-\frac{\omega}{s} t$, we obtain an effective Hamiltonian which can describe a solid state problem. Such a crystalline structure in $\Theta$ is reproduced in the time domain if we return to the laboratory frame, as the relation between $\Theta$ and $t$ is linear. Thus, if we locate a detector close to a classical resonant trajectory, the probability of detection of a particle as a function of time reproduces crystalline structure described by means of $H_{\rm eff}$ in the moving frame. 

In Refs.~\cite{Flicker2017,Flicker2017long} it was proven that stable orbits of classical dissipative systems can reveal quasi-crystal tiling in time. We will show that quantum properties of quasi-crystals in time can be investigated, see also \cite{Li2012,Huang2018}. In condensed matter physics quasi-crystals are systems which do not have any minimal part which appears periodically in space. Nevertheless, two or more unit cells are not placed randomly because a $d$-dimensional quasi-crystal can be constructed as a slice through a $2d$-dimensional periodic crystal \cite{Socolar1986,Janot1994,Senechal1995}. We will focus on the $d=1$ case when 1D quasi-crystal structure can be constructed as a cut through a 2D square lattice.
The cut with the line whose gradient is the golden ratio generates the Fibonacci quasi-crystal which can be also constructed with the help of the so-called inflation rule \cite{Flicker2017}: $B\rightarrow BS$ and $S\rightarrow B$ where $B$ and $S$ denote, e.g., big and small wells, respectively, of a potential energy of a single particle. Successive application of the inflation rule shows the process of growing of the quasi-crystal, i.e. $B\rightarrow BS\rightarrow BSB\rightarrow BSBBS\rightarrow BSBBSBSB\rightarrow \dots$. 

\begin{figure} 	            
\includegraphics[width=0.45\columnwidth]{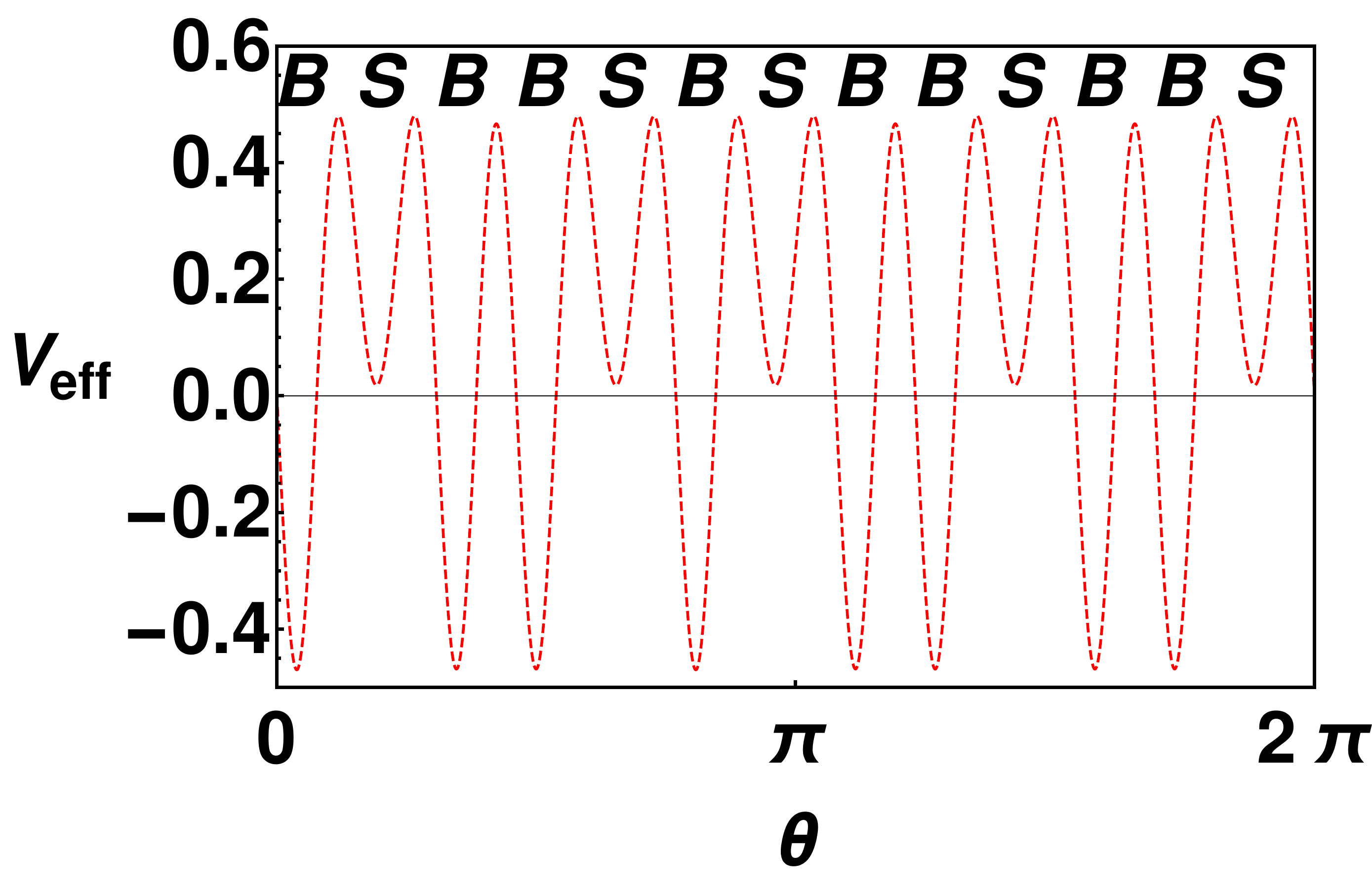}       
\hspace{0.02\columnwidth}
\includegraphics[width=0.45\columnwidth]{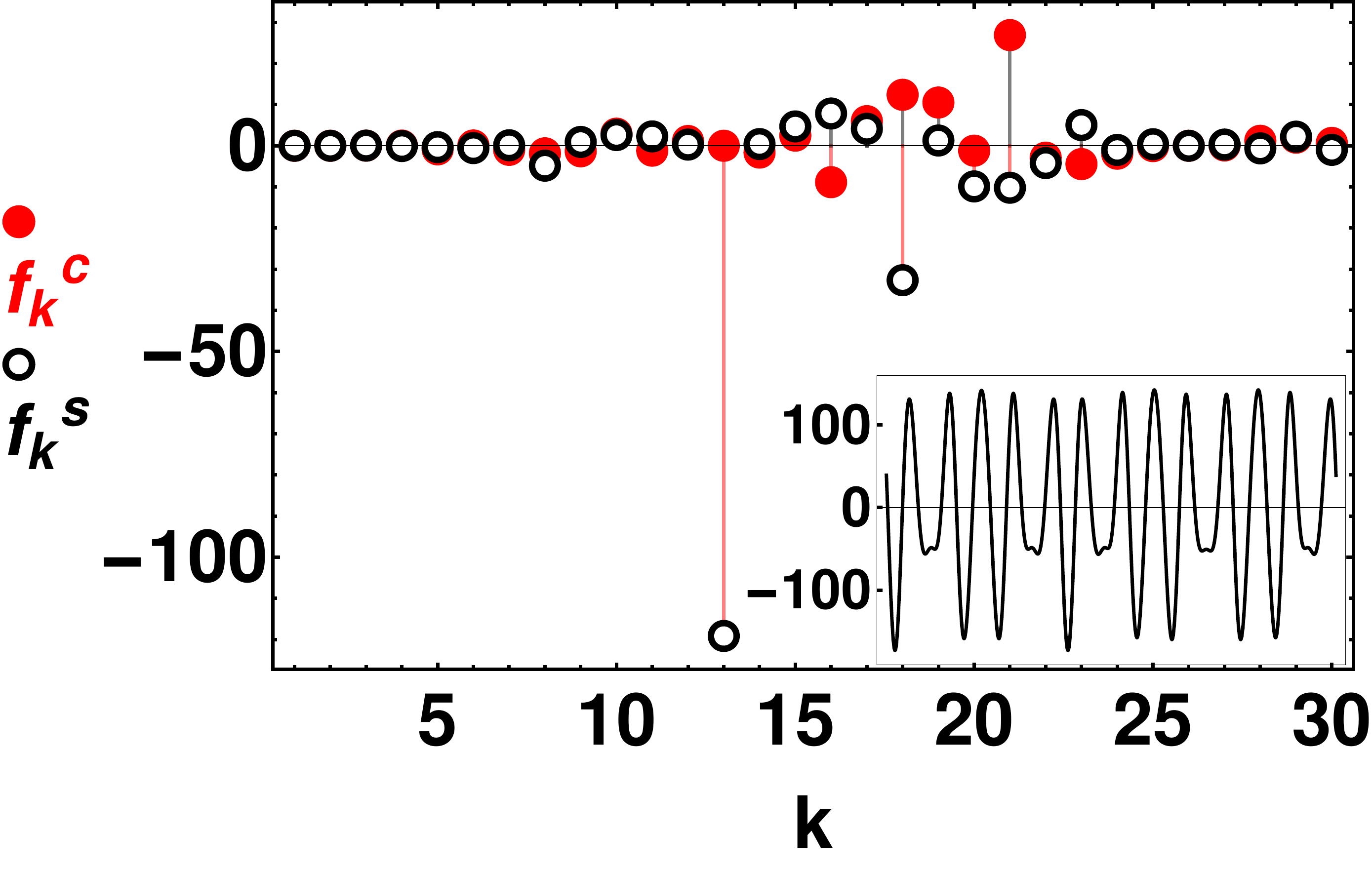}       
\caption{Time quasi-crystal: a particle bouncing on a vibrating mirror in a gravitational field in a 1D model. The 1:1 resonance condition is assumed.
Left: the effective potential $V_{\rm eff}(\Theta)$ with the quasi-crystal structure corresponding to the eighth Fibonacci number. Right: Fourier components of the periodic vibration of the mirror, $\frac{f(t)}{\omega^2}=\sum_k(f_k^c\cos k\omega t+f_k^s\sin k\omega t)$, that result in $V_{\rm eff}(\Theta)$ presented in the left panel. Full symbols are related to $f_k^c$, open symbols to $f_k^s$. The inset of this panel shows $\frac{f(t)}{\omega^2}$ over one period.}
\label{1Dquasi}   
\end{figure} 

In order to illustrate how to realize the Fibonacci quasi-crystal in the time domain experimentally, let us consider e.g., a particle which bounces on a vibrating mirror in the presence of a gravitational field \cite{Steane95,Lau1999,Bongs1999,Buchleitner2002} in a 1D model.
In the coordinate frame vibrating with the mirror, the mirror is fixed but the gravitation strength oscillates in time. Then the Hamiltonian of the system, in gravitational units \cite{Buchleitner2002}, reads $H=\frac{p^2}{2}+x+\lambda xf(t)$ where $f(t)=\sum_kf_ke^{ik\omega t}$ and $\frac{\lambda}{\omega^2}$ is related to the amplitude of the mirror vibration. The secular approximation leads to the previously derived effective Hamiltonian with $m_{\rm eff}=-\frac{\pi^2}{\omega^4}$ and $h_{n}=-\frac{(-1)^{n}}{n^2\omega^2}$ if the 1:1 resonance condition ($s=1$) is fulfilled. A proper choice of $f(t)$ allows one to realize the effective potential $V_{\rm eff}(\Theta)$ that reproduces any finite Fibonacci quasi-crystal. In Fig.~\ref{1Dquasi} we show what kind of driving leads to a quasi-crystal with the total number of big and small potential wells given by the seventh Fibonacci number. Transport properties in the quasi-crystal that can be analysed with the help of the effective Hamiltonian in the frame moving along the 1:1 resonant orbit will be observed in the time domain in the laboratory frame.  

\begin{figure} 	            
\includegraphics[width=0.45\columnwidth]{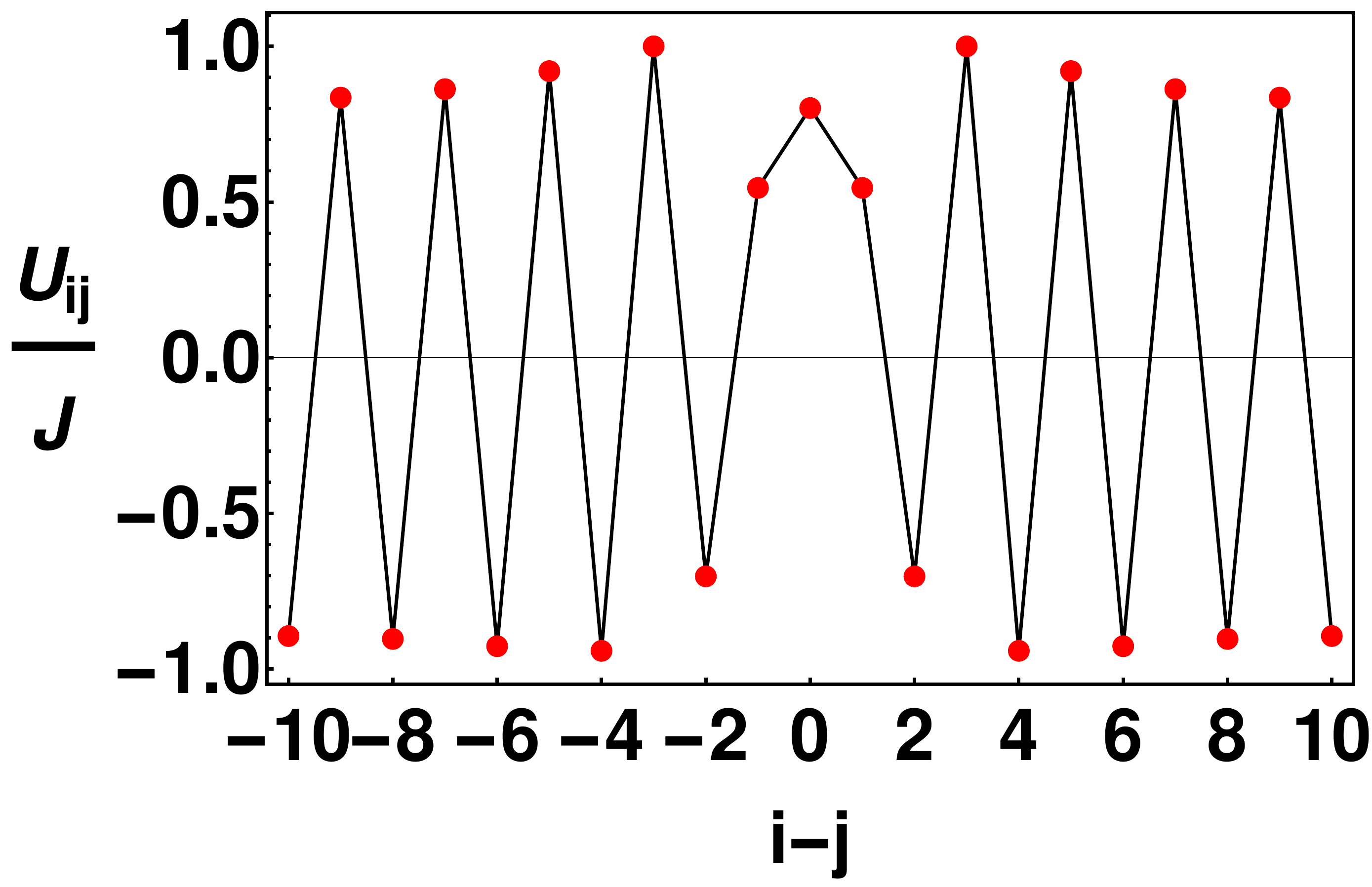}       
\includegraphics[width=0.45\columnwidth]{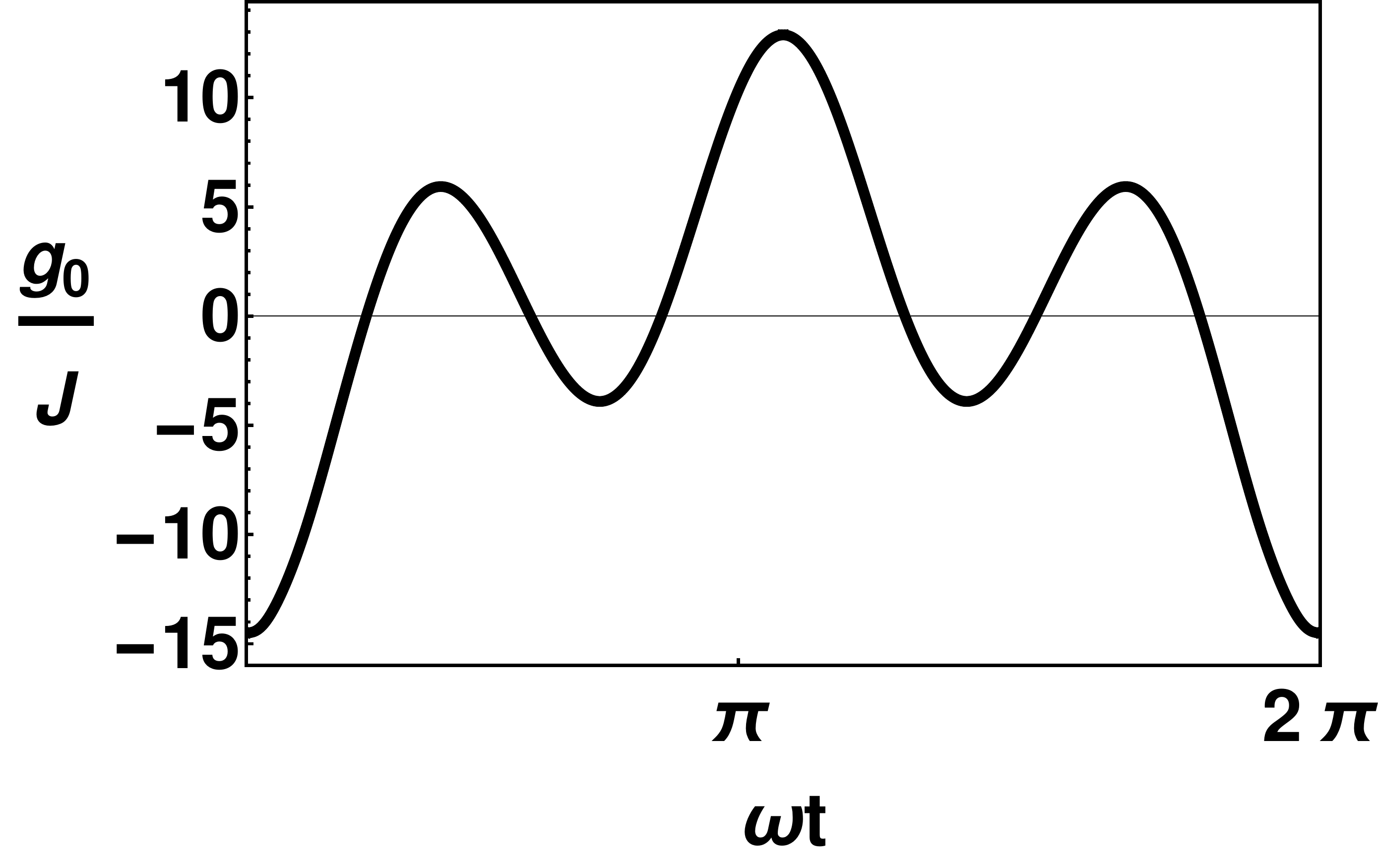}       
\caption{System with exotic interactions: ultra-cold atoms bouncing on a harmonically oscillating mirror in a 1D model. The 20:1 resonance condition is fulfilled and the many-body system is described by the Hamiltonian (\ref{bhh}). Left panel shows the interaction coefficients $U_{ij}$ corresponding to the scattering length $g_0(t)$ that is presented in the right panel. The frequency $\omega=2.8$ of the mirror oscillations and $\lambda=0.1$ which result in $J=3.7\times 10^{-5}$ and the gap of $302J$ between the lowest and first excited energy bands. Temporary interaction coefficients $\frac{s2\pi}{\omega}|g_0(t)|u_{ij}(t)\le 85J$.}
\label{many-body}   
\end{figure} 

Now we will demonstrate that periodically driven many-body systems allows for realization of solid state problems with exotic interactions. Let us illustrate this idea with ultra-cold atoms bouncing on a mirror which oscillates harmonically with frequency $\omega$. If the $s$:1 resonance condition is fulfilled, the single-particle effective Hamiltonian in the moving frame reads $H_{\rm eff}=\frac{P^2}{2m_{\rm eff}}+V_0\cos(s\Theta)$ where $m_{\rm eff}=-\frac{\pi^2s^4}{\omega^4}$ and $V_0=-\frac{\lambda(-1)^{s}}{\omega^2}$. 
Let us assume $s\gg 1$ and $V_0$ sufficiently big so that in the quantum description eigenvalues of ${H}_{\rm eff}$ form well separated energy bands and eigenstates are Bloch waves $e^{ik\Theta}v_k(\Theta)$ where $v_k(\Theta+\frac{2\pi}{s})=v_k(\Theta)$. Note that for a fixed position in the laboratory frame, $\theta=$const, the periodic Bloch waves character emerges in time, $e^{ik(\theta-\omega t/s)}v_k(\theta-\frac{\omega t}{s})$, with the period $\frac{2\pi}{\omega}$.
Due to the negative effective mass $m_{\rm eff}$, the effective Hamiltonian $H_{\rm eff}$ is bounded from above, not from below. Therefore, the first energy band possesses the highest energy. For simplicity, let us restrict ourselves to the first band and choose as the basis in the corresponding Hilbert subspace, the Wannier states $w_j=w(\Theta-j\frac{2\pi}{s})$ where $j$ denotes at which site of the effective potential a Wannier function is localized \cite{Dutta2015}. In the laboratory frame the Wannier states $w_j(x,t)$ describe localized wavepackets moving along the resonant trajectory. We assume the normalization $\int_0^{s2\pi/\omega}dt\langle w_j|w_j\rangle=1$. Thus, $s$ sites of the effective potential in the moving frame correspond to $s$ Wannier wavepackets evolving in the laboratory frame. The width of the first energy band of $H_{\rm eff}$ is determined by $J=-2\int_0^{s2\pi/\omega}dt\langle w_{i+1}|H_{\rm eff}|w_{i}\rangle$ which is an amplitude of nearest neighbour tunnelings.

In ultra-cold atomic gases interactions are described by the contact Dirac-delta potential, $g_0\delta(x)$, where $g_0$ is determined by the atomic scattering length which can be modulated in time by means of a Feshbach resonance \cite{Chin2010}. We will see that these contact interactions between atoms can result in exotic long-range interactions in the effective description of the resonantly driven many-body system (effective long-range interactions in the phase space crystals \cite{Guo2013,Guo2016} have been considered in \cite{Guo2016a,Liang2017}, see also \cite{Anisimovas2015}). For example in the case of bosonic particles, when we restrict ourselves to the Hilbert subspace spanned by Fock states $|\dots,n_{j},\dots\rangle$, where $n_{j}$ is the number of atoms occupying a mode $w_j$, we obtain a many-body effective Hamiltonian of the Bose-Hubbard form,
\be
\hat H_{\rm eff}=-\frac{J}{2}\sum_{\langle i, j\rangle}\hat a_{i}^\dagger\hat a_{j}+\frac12\sum_{i,j}U_{ij}\;\hat a_{i}^\dagger\hat a_{j}^\dagger\hat a_{j}\hat a_{i},
\label{bhh}
\ee
where the bosonic operators $\hat a_j$ annihilate particles in modes $w_j$'s and $U_{ij}=\int_0^{s2\pi/\omega}dtg_0(t)u_{ij}(t)$ with $u_{ij}(t)=2\int_0^\infty dx|w_i|^2|w_j|^2$ for $i\ne j$ and $u_{ii}=\int_0^\infty dx|w_i|^4$ \cite{Sacha15a}, where we assume that the atomic scattering length $g_0(t)$ can be modulated in time. The Hamiltonian (\ref{bhh}) is valid provided the interaction energy per particle is always smaller than the energy gap between the lowest and first excited energy bands of the single-particle system. A given interaction coefficient $U_{ij}$ is determined mostly by $g_0(t)$ at the moment when the corresponding wavepackets overlap. Suitable modulation of the scattering length $g_0(t)$ allows us to shape the interactions in (\ref{bhh}).
In order to perform a systematic analysis one can apply the singular value decomposition of the matrix $u_{ij}(t)$ where $(i,j)$ and $t$ are treated as indices of rows and columns, respectively. Left singular vectors tell us which sets of interaction coefficients $U_{ij}$ can be realized, while the corresponding right singular vectors give the recipes for $g_0(t)$. In Fig.~\ref{many-body} we present an example of the interaction coefficients and the corresponding function $g_0(t)$. In this example the magnitude of the interactions of a particle located at a given site with other particles located at the same or distant sites is nearly the same, but their repulsive or attractive character changes in an oscillatory way.

\begin{figure} 	            
\includegraphics[width=0.5\columnwidth]{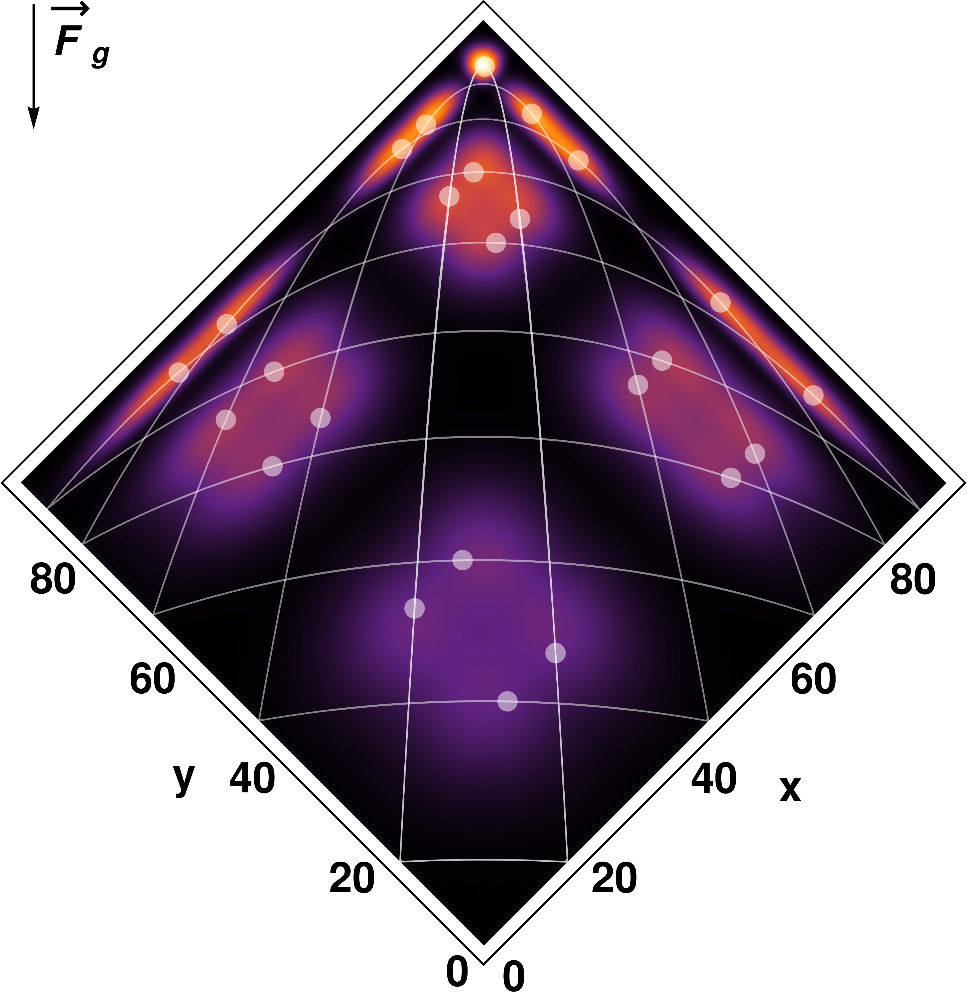}       
\hspace{0.02\columnwidth}
\includegraphics[width=0.45\columnwidth]{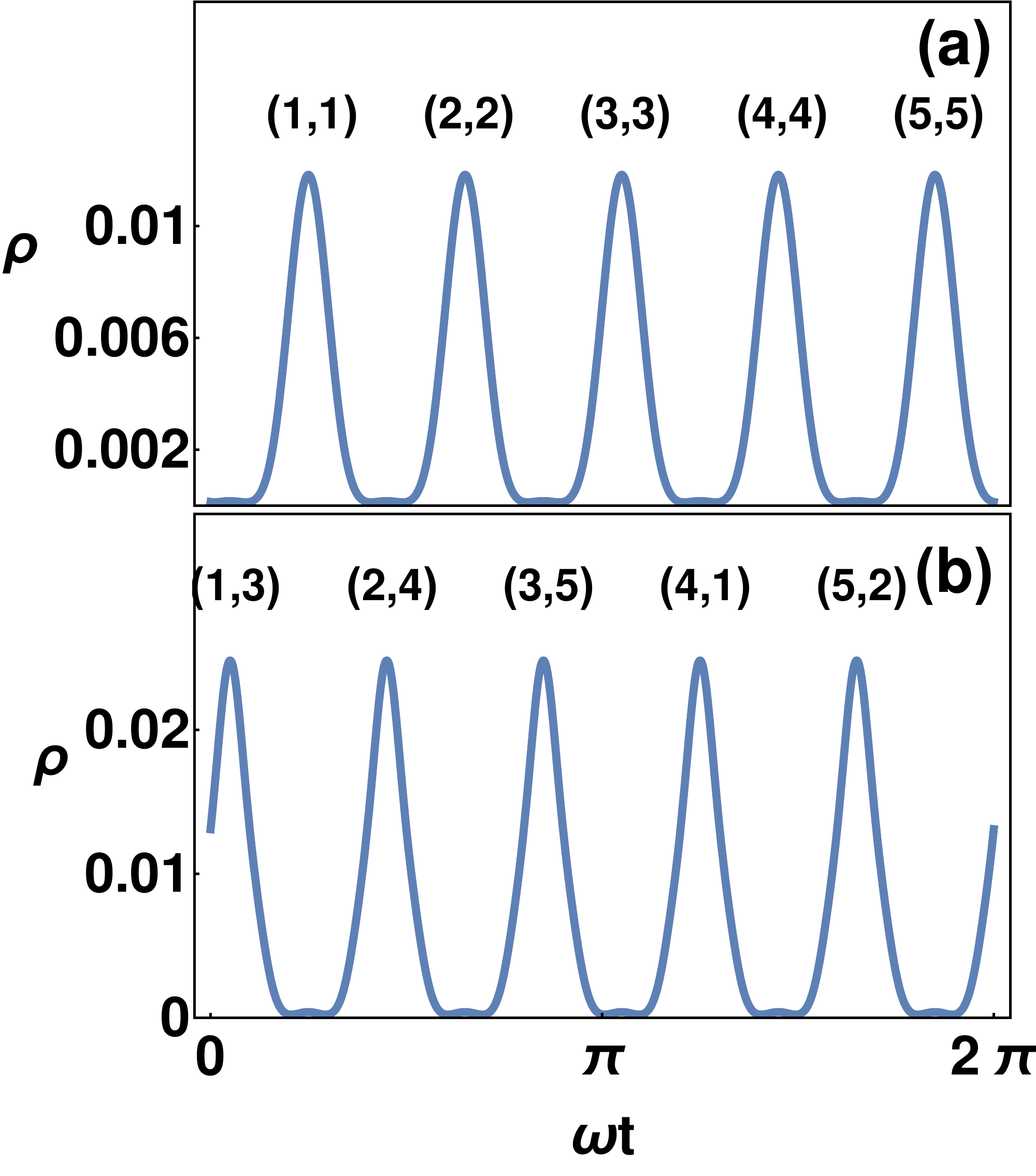}       
\caption{Time crystal with properties of a 2D space crystal: ultra-cold atoms bouncing between two perpendicular and harmonically oscillating mirrors with $\omega=1.05$, $\lambda=0.02$ and $\varphi=\frac{\pi}{2}$. The 5:1 resonance conditions are fulfilled and the many-body system is described by a 2D version of the Bose-Hubbard Hamiltonian (\ref{bhh}). For $g_0(t)=$const., the on-site interactions are dominant, i.e. $U_{ii}/g_0\in[0.2,0.3]J$, where different values correspond to different classical trajectories, while $U_{ij\ne i}/g_0<0.07J$. Left panel shows 25 Wannier wavepackets, i.e. $\rho(x,y,t)=\sum_{\vect j}|W_{\vect j}(x,y,t)|^2$, at $t=\frac{4}{\omega}$ and trajectories along which they propagate --- dots indicate positions of the centers of the wavepackets. Two perpendicular mirrors are located at $x=0$ and $y=0$ and they form $\frac{\pi}{4}$ angle with respect to the gravitational force $\vec F_g$. Right panels present $\rho(x,y,t)$ at $(x,y)=(90,90)$ (a) and $(82,82)$ (b) versus $t$ --- these plots reflect cuts of a square lattice described by the Bose-Hubbard model in the moving frame. Numbers in parentheses indicate which lattice sites $\vect j=(j_x,j_y)$ are located along the cuts.}
\label{2Dbouncer}   
\end{figure} 

Time is a single degree of freedom and it is hard to imagine multidimensional time crystals. However, we will see that resonantly driven systems can reveal properties of 2D or 3D space crystals in the time domain. Let us begin with a single particle bouncing between two mirrors that oscillate harmonically in two orthogonal directions with frequency $\omega$, see Fig.~\ref{2Dbouncer} --- generalization to the 3D case is straightforward. The single particle Hamiltonian reads $H=\frac{p_x^2+p_y^2}{2}+x+y+\lambda x\cos\omega t+\lambda y\cos(\omega t+\varphi)$ where $\varphi$ is the relative phase of the mirror oscillations. Assuming that for each of the two independent degrees of freedom the $s$:1 resonance condition is fulfilled we obtain (in terms of the action-angle variables and in the moving frame $\Theta_{j=x,y}=\theta_j-\frac{\omega}{s}t$) the effective Hamiltonian, $H_{\rm eff}=\frac{P_x^2+P_y^2}{2m_{\rm eff}}+V_0\left[\cos(s\Theta_x)+\cos(s\Theta_y)\right]$, which describes a particle in a 2D square lattice. For $s\gg 1$, eigenstates of $H_{\rm eff}$ are Bloch waves $e^{i(k_x\Theta_x+k_y\Theta_y)}v_{k_x}(\Theta_x)v_{k_y}(\Theta_y)$. When we a fix position in the laboratory frame, i.e. we fix $\theta_x$ and $\theta_y$, periodic character of Bloch waves emerges in time, $e^{i(k_x\theta_x+k_y\theta_y-(k_x+k_y)\omega t/s)}v_{k_x}(\theta_x-\frac{\omega t}{s})v_{k_y}(\theta_y-\frac{\omega t}{s})$. Different fixed values of $\theta_x$ and $\theta_y$ allows us to observe in the time domain different cuts of the square lattice described by $H_{\rm eff}$. We restrict ourselves to the first energy band of $H_{\rm eff}$ and define the Wannier state basis $W_{\vect j}=w_x(\Theta_x-j_x\frac{2\pi}{s})w_y(\Theta_y-j_y\frac{2\pi}{s})$ where $\vect j=(j_x,j_y)$ denotes at which site of the effective potential a Wannier function is localized. In the laboratory frame the Wannier states $W_{\vect j}(x,y,t)$ describe localized wavepackets moving along resonant trajectories. The shape of the trajectories depends on the relative phase $\varphi$ of the mirrors oscillations. For $\varphi\ne \frac{\pi}{2}$ there are $s$ different trajectories in the configuration space and $s$ wavepackets $W_{\vect j}(x,y,t)$ moving along each of them, see Fig.~\ref{2Dbouncer}. Thus, $s^2$ sites of the effective potential in the moving frame correspond to $s^2$ Wannier wavepackets evolving in the laboratory frame. Switching to the many-body case we obtain, for ultra-cold bosons, a 2D version of the Hamiltonian (\ref{bhh}). If the scattering length is not modulated in time, i.e. $g_0(t)=$const., the on-site interactions are dominant and the system reproduces, in the moving frame, a 2D squared lattice problem with on-site interactions \cite{Dutta2015}. If we locate detectors at different positions in the laboratory frame, the time dependence of the probabilities of detection reflects cuts of the 2D square lattice, see Fig.~\ref{2Dbouncer}.

\begin{figure} 	            
\includegraphics[width=0.4\columnwidth]{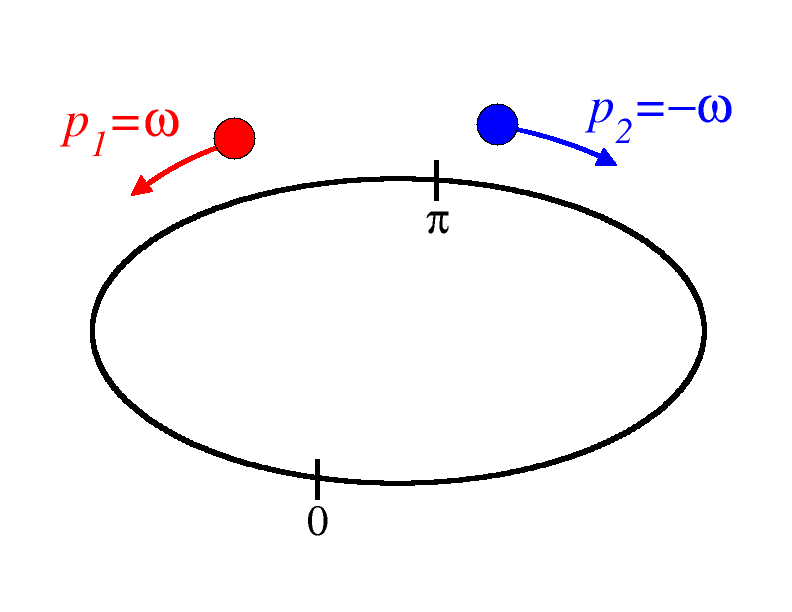}       
\includegraphics[width=0.45\columnwidth]{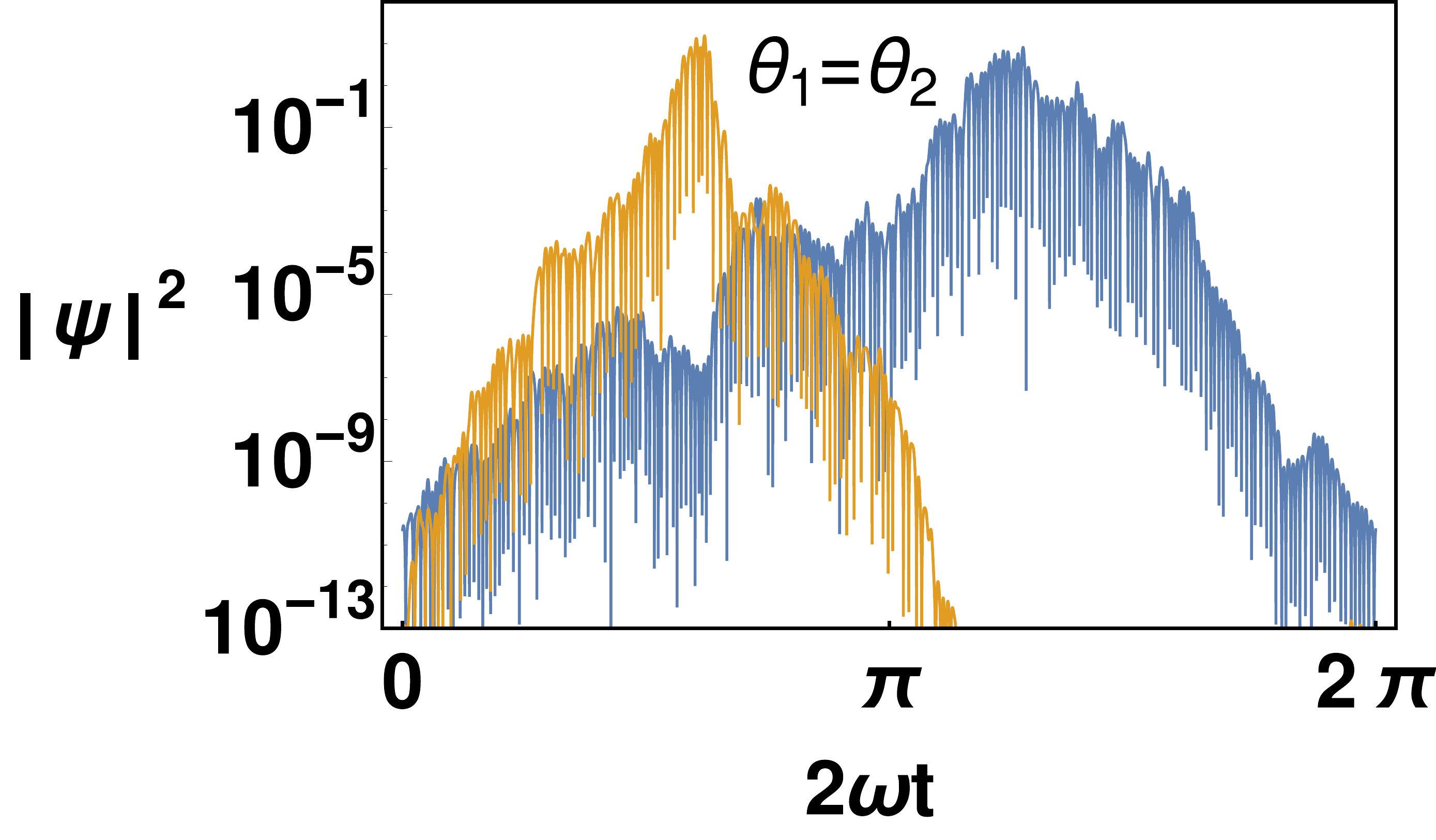}       
\caption{Two atoms bound together due to destructive interference. Left panel shows schematic plot of an experiment where two distinguishable atoms move on a ring in opposite directions with momenta $\pm\omega$. Resonant modulation of atomic scattering length allows one to create a molecule where the atoms are bound together due to destructive interference, i.e. in the moving frame eigenstates of the system are Anderson localized. Right panel presents probability densities for detection of the atoms at $\theta_1=\theta_2$ in the laboratory frame versus $t$ for two eigenstates related to energies $E=6020$ (orange line) and 10460 (blue line) for $\lambda=5660$ and $k_0=500$.}
\label{2Ddetection}   
\end{figure} 

Finally, let us show that periodic driving allows one to create a molecule where Anderson localization is responsible for the binding of two atoms. Assume that two atoms move on a ring and their scattering length is modulated in time employing a Feshbach resonance so that the Hamiltonian of the system reads $H=\frac{p_1^2+p_2^2}{2}+2\pi \lambda\delta(\theta_1-\theta_2)f(t)$ where $\lambda$ is a constant, $f(t)=\sum_{k\ne 0}f_ke^{ik\omega t}$ and $\theta_{1,2}$ denote positions of the atoms on the ring. If the first atom is moving in the clockwise direction with momentum $p_1\approx \omega$, and the other in the anticlockwise direction with $p_2\approx -\omega$, then the secular approximation results in $H_{\rm eff}=\frac{P_1^2+P_2^2}{2}+\lambda V_{\rm eff}(\Theta_1-\Theta_2)$ in the moving frame, i.e. $\Theta_1=\theta_1-\omega t$ and $\Theta_2=\theta_2+\omega t$. Interactions between atoms are described by the effective potential $V_{\rm eff}=\sum_nf_{-2n}e^{in(\Theta_1-\Theta_2)}$ whose shape can be engineered at will by a suitable choice of the Fourier components $f_k$ of the periodic driving. For example if $f_k=\frac{1}{\sqrt{k_0}}e^{i\varphi_k}$ for $|k|\le \frac{k_0}{2}$ and zero otherwise, where $\varphi_{k}=-\varphi_{-k}$ are random variables chosen from a uniform distribution, the atoms interact via the effective disordered potential characterized by the correlation length $\frac{\sqrt{2}}{k_0}$ and the standard deviation $\lambda$. Then, eigenstates $\psi(\Theta_1-\Theta_2)$ of $H_{\rm eff}$ are Anderson localized around different values $\theta_0$ of the relative coordinate \cite{MuellerDelande:Houches:2009}, i.e. $|\psi|^2\propto e^{-|\Theta_1-\Theta_2-\theta_0|/l_0}$, provided the localization length $l_0\ll 2\pi$ --- within the Born approximation $l_0=\frac{Ek_0^2}{\pi \lambda^2}$, which is valid when $\frac{\lambda^2}{k_0^2}\ll E\ll \frac{k_0^2}{4}$, where $E$ is energy of the system in the moving frame \cite{Kuhn:Speckle:NJP07,Giergiel2017}. Hence, we are dealing with a situation where two atoms are bound together not by attractive interactions but due to destructive interference, i.e. due to Anderson localization phenomenon induced by disordered mutual interactions \cite{supplement}. 
If atoms are identical bosons (fermions), an eigenstate must be symmetric (antisymmetric) under their exchange. This symmetry is easily restored because we can exchange the role of the atoms. That is, the first atom can move in the anticlockwise direction, $p_1\approx -\omega$, while the other one in the clockwise direction, $p_2\approx \omega$. Consequently, proper Floquet eigenstates for bosons or fermions, in the laboratory frame, read $\psi(\theta_1-\theta_2-2\omega t)\pm\psi(\theta_2-\theta_1-2\omega t)$. 
Experimental demonstration of two atoms bound due to destructive interference seems straightforward if atoms are prepared in a toroidal trap \cite{Clifford1998,Wright2000,Wang2009}. 

In summary, we have shown that a wide class of condensed matter problems can be realized in the time domain if single-particle or many-body systems are resonantly driven. It opens up unexplored territory for investigation of condensed matter physics in time and for the invention of novel {\it time devices} because time is our new ally. As an example we have demonstrated that periodic driving allows one to realize molecules where atoms are bound together not due to attractive mutual interactions but due to destructive interference.

Support of the National Science Centre, Poland via Projects No. 2016/20/W/ST4/00314 (K.G.) and No. 2016/21/B/ST2/01095 (K.S.) is acknowledged. This work was performed with the support of EU via Horizon2020 FET project QUIC (No. 641122).




\section{Supplemental Material}

In this Supplemental Material we present an analysis of the validity of the effective Hamiltonian approach for two examples: a particle bouncing on a vibrating mirror and formation of a molecule where two atoms are bound together due to destructive interference. However, we begin with a short introduction to the classical and quantum description of periodically driven systems. 

If the Hamiltonian of, e.g., a single-particle system in one-dimensional space depends explicitly on time, $H(x,p,t)$, the energy is not conserved. In the classical description one can extend the phase space of the system by defining $t$ as an additional {\it dimension} of the configuration space and $p_t=-H$ as the conjugate momentum \cite{Lichtenberg1992s,Buchleitner2002s}. Then, the new Hamiltonian in such an extended phase space reads ${\cal H}=H(x,p,t)+p_t$ and the motion of a particle is parametrized by some fictitious time $\tau$. Because $\frac{dt}{d\tau}=\frac{\partial {\cal H}}{\partial p_t}=1$, $t$ and $\tau$ are essentially identical. The new Hamiltonian does not depend explicitly on $\tau$ and thus it is conserved as $\tau$ evolves.

In the quantum description there are no energy eigenstates because the energy is not conserved. However, if a particle is periodically driven, $H(x,p,t+\frac{2\pi}{\omega})=H(x,p,t)$, we can look for a kind of stationary states of the form of $\psi_n(x,t)=e^{-iE_nt}v_n(x,t)$ where $v_n(x,t+\frac{2\pi}{\omega})=v_n(x,t)$. Indeed, substituting $\psi_n(x,t)$ in the time-dependent Schr\"odinger equation we obtain the eigenvalue problem,
\be
\left(H(x,p,t)-i\partial_t\right)v_n(x,t)=E_nv_n(x,t),
\ee
where $v_n(x,t)$ fulfills periodic boundary conditions in time \cite{Shirley1965s,Buchleitner2002s}. 
Thus, one may define a new Hamiltonian (so-called Floquet Hamiltonian) $H_F=H-i\partial_t$ whose eigenstates (called Floquet states) are time-periodic and form a complete basis in the Hilbert space of the system at any time $t$ \cite{Shirley1965s}. The Floquet Hamiltonian can be considered as the quantized version of the classical Hamiltonian $\cal H$ in the extended phase space where $p_t\rightarrow-i\partial_t$. 

In the Letter we show different periodically driven systems whose Floquet states possess properties of condensed matter systems in the time domain. In order to identify systems with such properties and find suitable parameters we begin with the classical approach, obtain effective Hamiltonians and then switch to the quantum description.

\begin{figure*}[!ht] 	            
\includegraphics[width=0.49\textwidth]{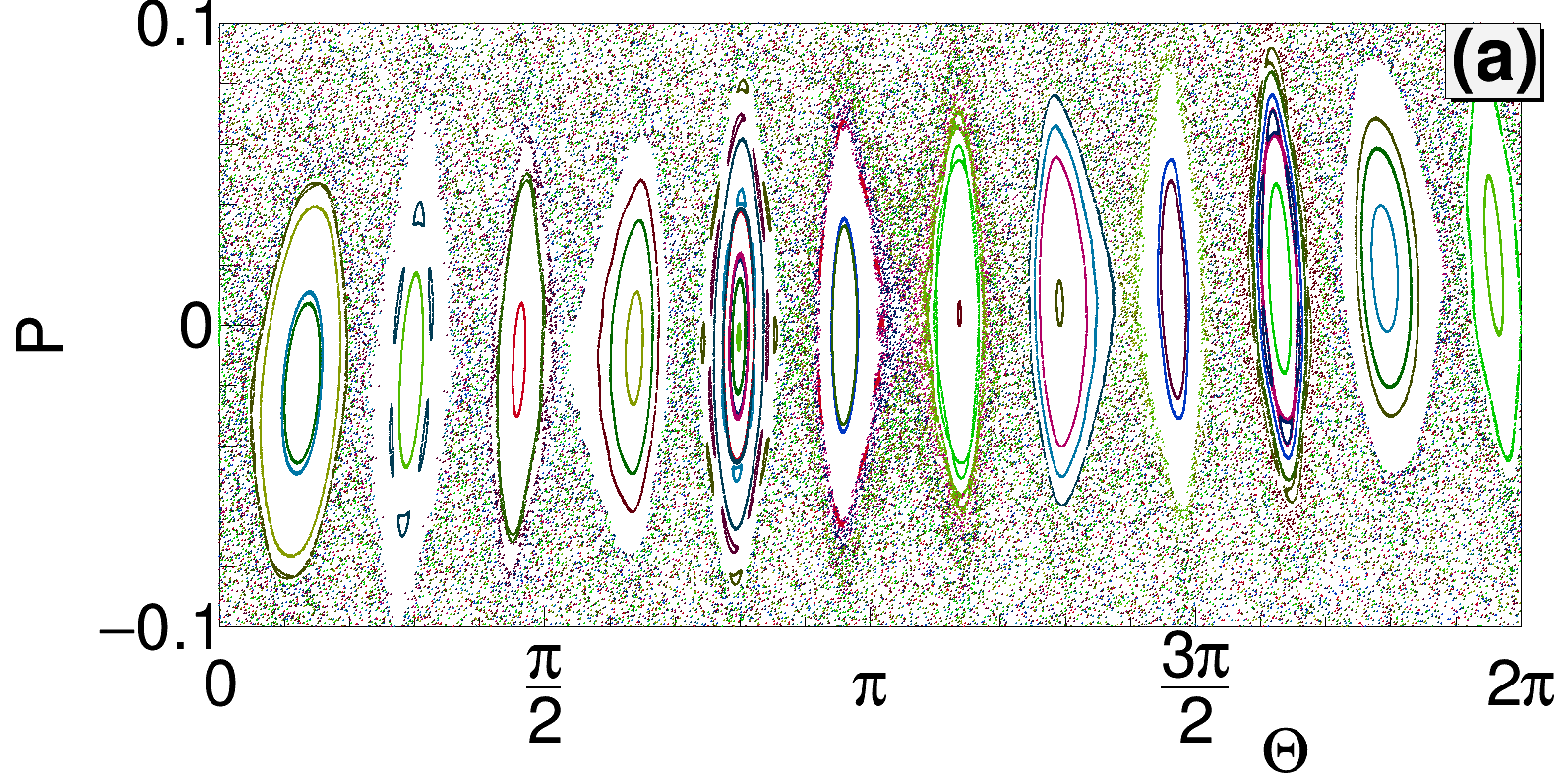}       
\includegraphics[width=0.49\textwidth]{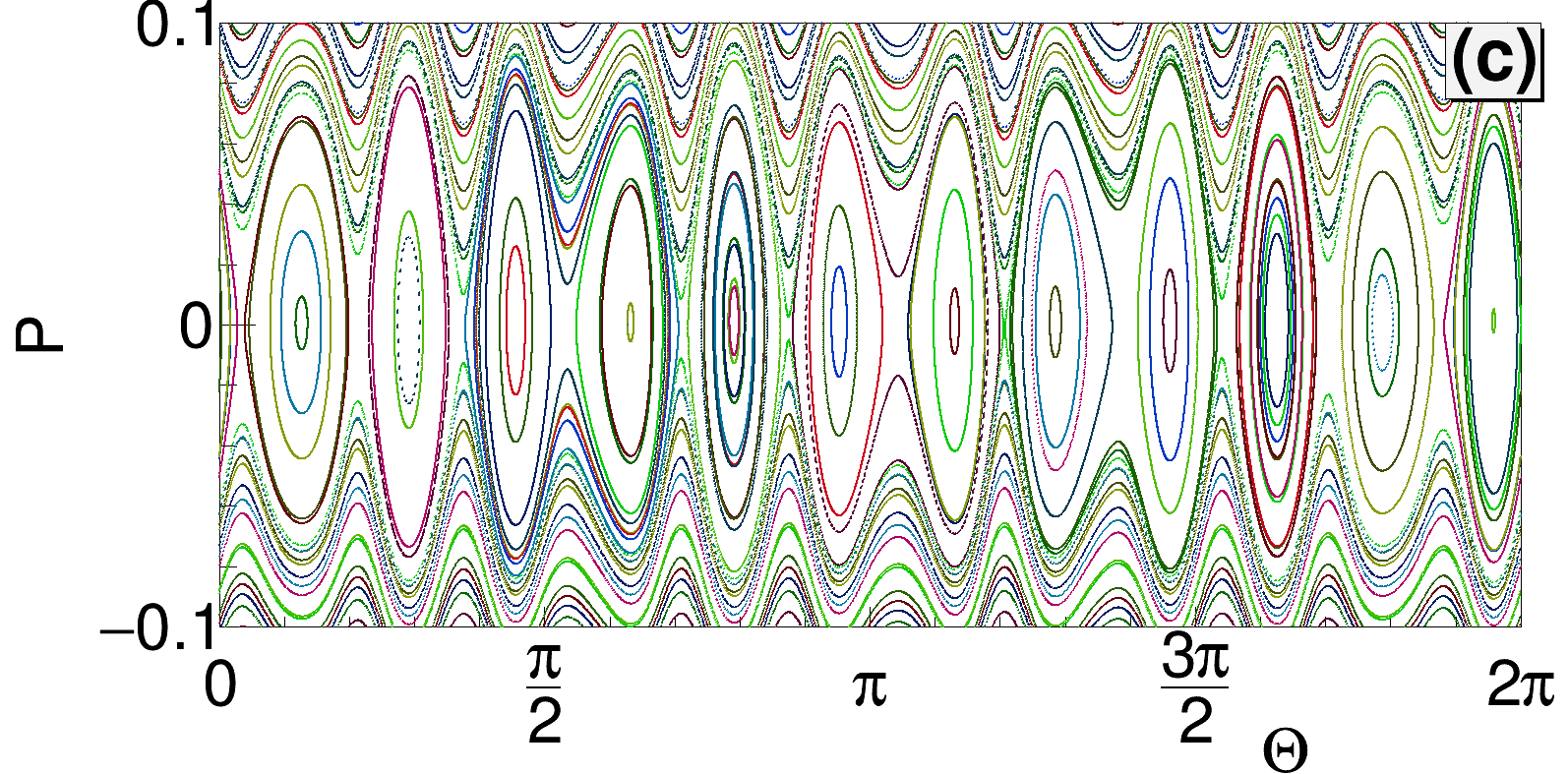}       
\includegraphics[width=0.49\textwidth]{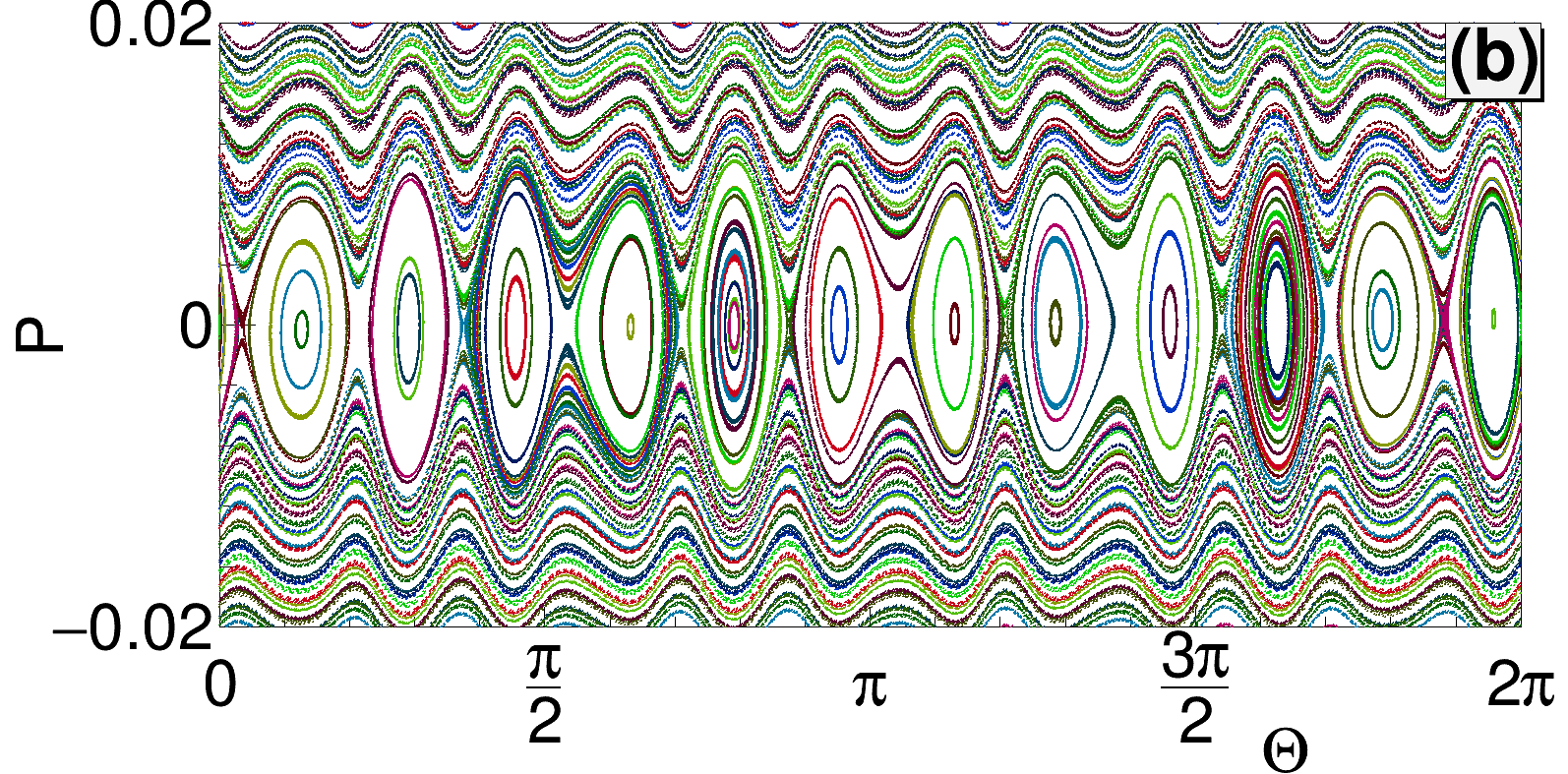}       
\includegraphics[width=0.49\textwidth]{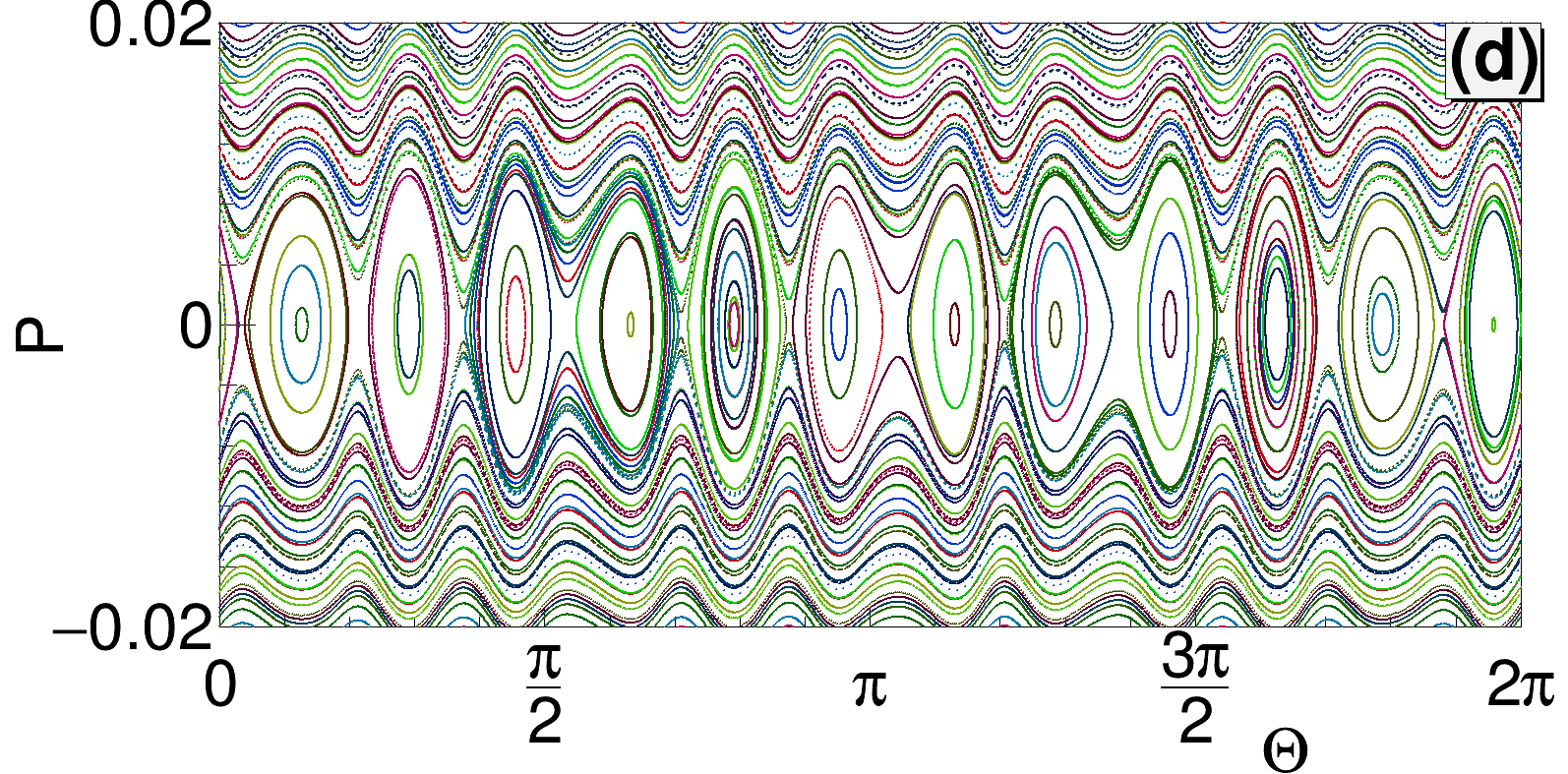}       
\caption{Panels (a) and (b): stroboscopic picture of the phase space of a particle bouncing on a vibrating mirror. Equations of motion generated by the full Hamiltonian (\ref{sh1}) have been integrated and after every period $\frac{2\pi}{\omega}$ of the mirror oscillations position of a particle in the $(\Theta,P)$ space has been plotted. The variable $\Theta=\theta-\omega t$ denotes position of a particle in the frame moving along the 1:1 resonant orbit. The canonically conjugate momentum $P=I-I_0$ where $I_0$ fulfills the 1:1 resonant condition $\Omega(I_0)=\omega$ (we have chosen such $\omega$ that $I_0=1$). The strength of the perturbation $\lambda=0.6$ (a) and $\lambda=0.01$ (b). In panels (c) and (d) the phase space portraits generated by the effective Hamiltonian (\ref{seffh}) are presented which correspond to the same parameters of the system as in the panels (a) and (b), respectively.}
\label{sfig1}   
\end{figure*} 

\subsection{Particle bouncing on a vibrating mirror} 
Let us consider a particle which bounces on a periodically vibrating mirror in the presence of the gravitational force \cite{Buchleitner2002s}. In the frame moving with the mirror, the mirror does not vibrate but the gravitational potential changes periodically in time. The Hamiltonian of the system, in gravitational units, reads $H=H_0+H_1$ with
\be
H_0(x,p)=\frac{p^2}{2}+x, \quad\quad H_1(x,t)=\lambda xf(t),
\label{sh1}
\ee
where $f(t+2\pi/\omega)=f(t)$ and $\lambda$ determines the strength of the perturbation. 

For $\lambda=0$, the classical description of the system becomes very simple if we apply the canonical transformation to the so-called action-angle variables $(I,\theta)$ \cite{Lichtenberg1992s}. Then, $H_0=H_0(I)=\frac12 \left(3\pi I\right)^{2/3}$ where the action $I$ (new momentum) is a constant of motion and the conjugate angle $\theta$ (new position variable) changes linearly in time. Now, $\theta(t)=\Omega t+\theta(0)$ where $\Omega(I)=\frac{dH_0(I)}{dI}$ is the frequency of a periodic motion of the unperturbed particle \cite{Buchleitner2002s}. The portrait of the $(\theta,I)$ phase space is very simple because it consists of straight lines corresponding to different values of $I=$constant. 

When $\lambda\ne 0$, the phase space structure changes around resonant values of $I$. Let us focus on $f(t)=\sum_{k}f_ke^{ik\omega t}$ where $f_k$'s correspond to the time quasi-crystal structure presented in Fig.~1 of the Letter. In the present Supplemental Material in Fig.~\ref{sfig1}(a)-\ref{sfig1}(b) we show a stroboscopic picture of the $(\Theta,P)$ space obtained by integration of the classical equations of motion generated by the Hamiltonian (\ref{sh1}) where $\Theta=\theta-\omega t$ and $P=I-I_0$ with $I_0$ corresponding to the 1:1 resonance condition between the driving force and the unperturbed particle motion, i.e. $\omega=\Omega(I_0)$. We have chosen $I_0=1$ and two different values of $\lambda$. If $\lambda$ is sufficiently small the phase space portrait around elliptical islands does not reveal chaotic motion and we may expect that the first order secular approximation \cite{Lichtenberg1992s} is able to perfectly describe motion of a particle close to the resonant value of $I$. The Hamiltonian (\ref{sh1}) in the frame moving along the resonant orbit, i.e. in the $\Theta=\theta-\omega t$ and $I$ variables, reads
\bea
H&=&H_0(I)-\omega I+\lambda f(t)\sum_n h_n(I)e^{in(\Theta+\omega t)},
\label{saah}
\eea
where $h_0=\left(\frac{\pi I}{\sqrt{3}}\right)^{2/3}$ and $h_n=\frac{(-1)^{n+1}}{n^2}\left(\frac{3I}{\pi^2}\right)^{2/3}$ for $n\ne 0$.
In the moving frame, $\Theta$ and $I$ are slow variables if $P=I-I_0\approx 0$. Then, averaging (\ref{saah}) over fast {\it time variable} and performing Taylor expansion around the resonant value of $I$ lead to the effective secular Hamiltonian,
\be
H_{\rm eff}=\frac{P^2}{2m_{\rm eff}}+\lambda\sum_n h_{n}(I_0)f_{-n}e^{in\Theta},
\label{seffh}
\ee  
where $m_{\rm eff}=-\frac{\pi^2}{\omega^4}$.
The phase space portraits generated by (\ref{seffh}) are shown in Fig.~\ref{sfig1}(c)-\ref{sfig1}(d) for the values of $\lambda$ corresponding to the exact portraits presented in Fig.~\ref{sfig1}(a)-\ref{sfig1}(b). In the case when $\lambda=0.01$, the effective Hamiltonian results and the exact data are identical. Thus, the motion of a particle in the quasi-crystal potential predicted by the effective Hamiltonian is reproduced by the full classical dynamics provided $\lambda$ is sufficiently small.

In order to switch to quantum effective description we can either perform quantization of the classical effective Hamiltonian (\ref{seffh}) or apply quantum version of secular approximation for the Hamiltonian (\ref{sh1}) \cite{Berman1977s}. Let us first discuss the former approach. Classical equations of motion possess the scaling symmetry which implies that by a proper rescaling of the parameters and dynamical variables of the system we obtain the same behavior as presented in Fig.~\ref{sfig1} but around arbitrary value of $I_0\ne 1$ \cite{Buchleitner2002s}. That is, when we redefine $\omega'=I_0^{-1/3}\omega$ and $\lambda'=\lambda$ we can use the results presented in Fig.~\ref{sh1} if we rescale $p'=I_0^{1/3}p$, $x'=I_0^{2/3}x$ and $t'=I_0^{1/3}t$. In the quantum description the scaling symmetry is broken because the Planck constant sets a scale in the phase space,
\be
[x,p]=i \quad \Rightarrow \quad [x',p']=\frac{i}{I_0}.
\ee
For $I_0\gg 1$, the quantized version of the effective Hamiltonian (\ref{seffh}), i.e. when $P\rightarrow -i\frac{\partial}{\partial \Theta}$, provides perfect quantum description of the resonant behavior of the system. The same quantum results can be obtained by applying the quantum secular approach \cite{Berman1977s} which yields
\be
\la n'|H_{\rm eff}|n \ra=\left(E_n-n\omega\right)\delta_{nn'}+\lambda \la n'|x|n\ra f_{n-n'},
\label{sqeffh}
\ee
where $|n\ra$'s are eigenstates of the unperturbed ($\lambda=0$) system and $2^{1/3}E_n$ are zeros of the Airy function \cite{Buchleitner2002s}. Equation~(\ref{sqeffh}) has been obtained by switching to the moving frame, with the help of the unitary transformation $\hat U=e^{i\hat n\omega t}$, and by averaging the Hamiltonian over the short time scale $\frac{2\pi}{\omega}$ \cite{Berman1977s}.

\subsection{Molecule formation due to destructive interference}

\begin{figure} 	            
\includegraphics[width=0.9\columnwidth]{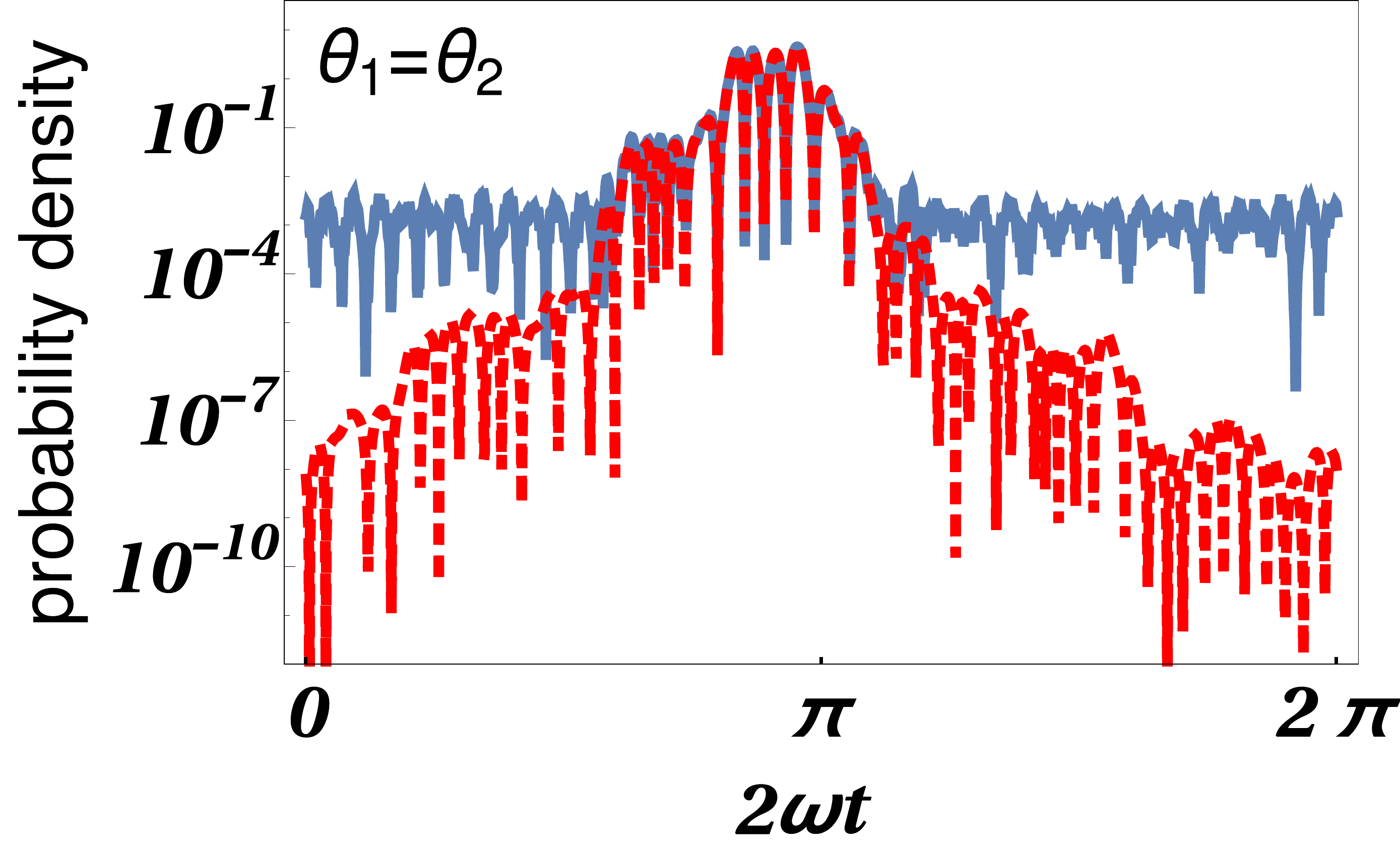}       
\includegraphics[width=0.9\columnwidth]{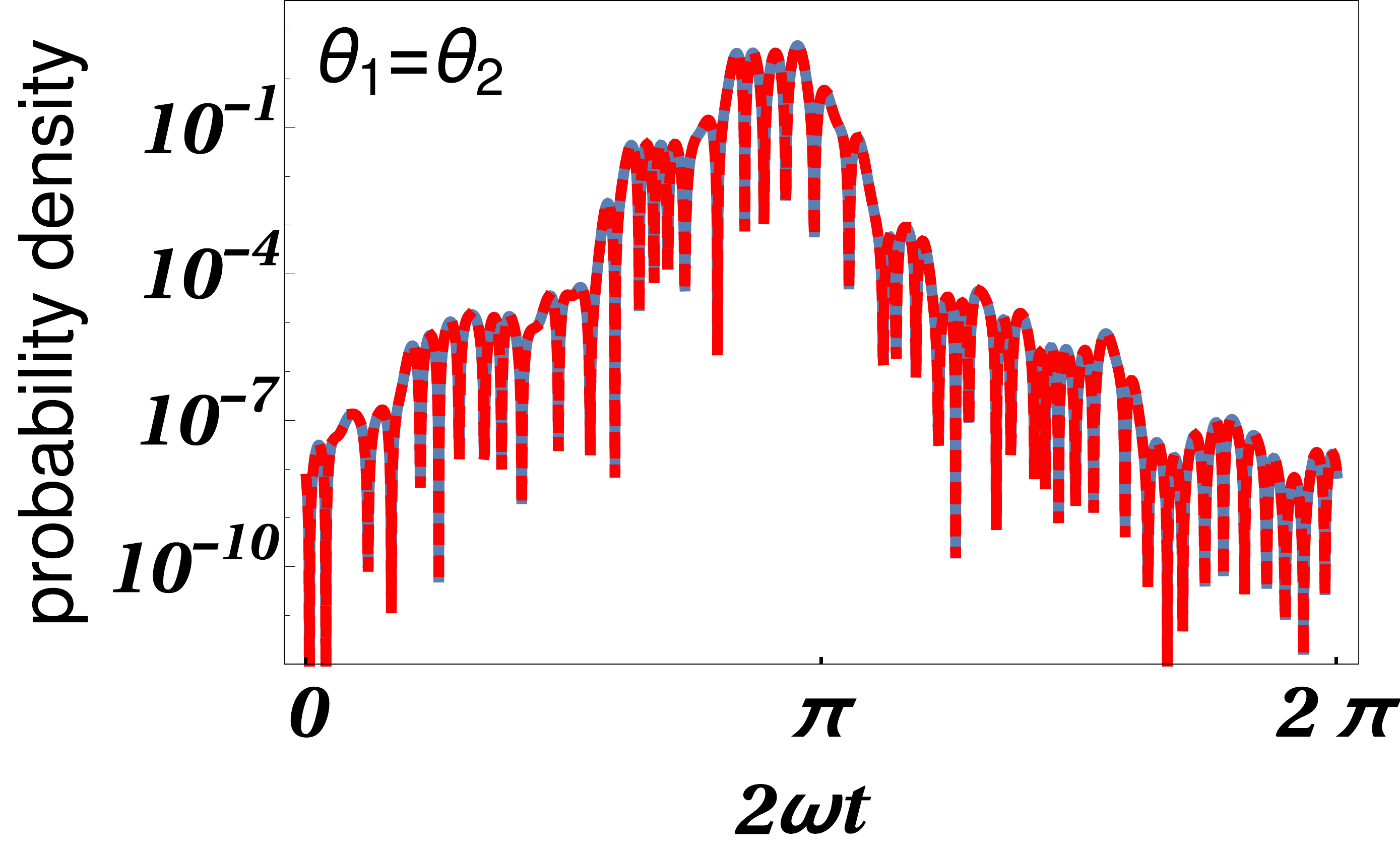}       
\caption{Red dashed lines show probability densities for detection of two atoms at $\theta_1=\theta_2$ in the laboratory frame versus $t$ for an eigenstate of the effective Hamiltonian (\ref{seffhm}) related to the energy $E=616$, $\lambda=600$ and $k_0=100$. Blue solid lines depict the probability density of the corresponding Floquet eigenstate of the full Hamiltonian (\ref{sfullhm}) for $\omega=10^4$ (top panel) and $\omega=3\times10^5$ (bottom panel).}
\label{sfig3}   
\end{figure} 

Let us consider two atoms which move on a ring and interact via a contact potential,
\be
H=\frac{p_1^2+p_2^2}{2}+2\pi\lambda \delta(\theta_1-\theta_2)f(t),
\label{sfullhm}
\ee
where $f(t)=\sum_{k\ne 0}f_ke^{i\omega t}$ with $f_k=\frac{1}{\sqrt{k_0}}e^{i\varphi_k}$ for $|k|\le \frac{k_0}{2}$ and zero otherwise. $\varphi_k=-\varphi_{-k}$ are random variables chosen from the uniform distribution. 

If the momenta of atoms fulfill $p_1\approx \omega$ and $p_2\approx -\omega$, then in the moving frame, 
\bea
\Theta_1=\theta_1-\omega t, && P_1=p_1-\omega, \cr 
\Theta_2=\theta_2+\omega t, && P_2=p_2+\omega, 
\eea
the secular approximation results in the effective Hamiltonian
\be
H_{\rm eff}=\frac{P_1^2+P_2^2}{2}+\lambda\sum_nf_{-2n}e^{in(\Theta_1-\Theta_2)},
\label{seffhm}
\ee
which describes two particles interacting via a disorder potential. For a suitable choice of the system parameters, two atoms can be bound together due to Anderson localization. 

The effective Hamiltonian does not depend on $\omega$ but its validity does. For a given $\lambda$, the effective Hamiltonian (\ref{seffhm}) is always valid if $\omega$ is sufficiently large as the second order corrections are proportional to $\frac{\lambda^2}{\omega^2}$.
This is illustrated in Fig.~\ref{sfig3} where we compare an eigenstate of (\ref{seffhm}) with the corresponding Floquet eigenstate of the full Hamiltonian (\ref{sfullhm}) for two different values of $\omega$. Thus, for sufficiently large $\omega$, higher order terms neglected in the first order secular approximation do not modify the Anderson localization phenomena predicted by the effective Hamiltonian (\ref{seffhm}).



\end{document}